# Nonparaxial Mathieu and Weber accelerating beams

Peng Zhang,[1] Yi Hu,[2,3] Tongcang Li,[1] Drake Cannan,[4] Xiaobo Yin,[1,5] Roberto Morandotti,[2] Zhigang Chen,[3,4] and Xiang Zhang[1,5]*

[1]*NSF Nanoscale Science and Engineering Center, 3112 Etcheverry Hall, University of California, Berkeley, CA 94720, USA*

[2]*Institut National de la Recherche Scientifique, Varennes, Québec J3X 1S2, Canada*

[3]*TEDA Applied Physics School, Nankai University, Tianjin 300457, China*

[4]*Department of Physics and Astronomy, San Francisco State University, San Francisco, California 94132, USA*

[5]*Materials Science Division, Lawrence Berkeley National Laboratory, 1 Cyclotron Road, Berkeley, CA 94720, USA*

* Corresponding author: xiang@berkeley.edu

We demonstrate both theoretically and experimentally nonparaxial Mathieu and Weber accelerating beams, generalizing the concept of previously found accelerating beams. We show that such beams bend into large angles along circular, elliptical or parabolic trajectories but still retain nondiffracting and self-healing capabilities. The circular nonparaxial accelerating beams can be considered as a special case of the Mathieu accelerating beams, while an Airy beam is only a special case of the Weber beams at the paraxial limit. Not only generalized nonparaxial accelerating beams open up many possibilities of beam engineering for applications, but the fundamental concept developed here can be applied to other linear wave systems in nature, ranging from electromagnetic and elastic waves to matter waves.

Self-accelerating beams have stimulated growing research interest since the concept of Airy wave packets was introduced from quantum mechanics [1] into optics in 2007 [2]. As exact solutions of the paraxial wave equation (which is equivalent to the Schrödinger equation), Airy beams propagate along parabolic trajectories and are endowed with nondiffracting and self-healing properties [2,3]. In the past five years, Airy beams have been studied intensively both theoretically and experimentally [4]. Possible applications of Airy beams have also been proposed and demonstrated, including guiding micro-particles [5], producing curved plasma channels [6], and dynamically routing surface plasmon polaritons [7]. However, it is important to point out that Airy beams are inherently subjected to the paraxial limit. As such, when a self-accelerating Airy beam moves along a parabola and eventually bends into a large angle, it will escape its domain of existence and finally break down. Recently, research efforts have been devoted to overcome the paraxial limit of Airy beams, and circular nonparaxial accelerating beams (NABs) have been identified theoretically and demonstrated experimentally [8-11]. Those NABs, found as exact solutions of the Helmholtz equation (HE) in circular coordinates, travel along circular trajectories beyond the paraxial limit, in a fundamentally different fashion from the well-known Bessel, Mathieu, and parabolic nondiffracting beams that travel along straight lines [12-14]. However, unlike Airy beams, the circular NABs cannot be simply scaled (by squeezing or stretching the transverse coordinates) to obtain different accelerations, but their beam paths have to be pre-designed to match only certain circular trajectories defined by the Bessel functions [10,11]. This naturally brings about a series of fundamental questions: Can a NAB bend itself along other trajectories rather than circles? If so, would such a beam maintain its nondiffracting and self-healing properties while bending to large angles and different paths? Is it possible to find a NAB that can be scaled to control the acceleration, thus leading to beneficial practical implementations?

In this Letter, we demonstrate both theoretically and experimentally nonparaxial Mathieu accelerating beams (MABs) and Weber accelerating beams (WABs), generalizing the concept of previously discovered accelerating beams into the full domain of the wave equation. Such new families of accelerating beams, found as exact solutions of the HE in different coordinate systems without the need of using the paraxial approximation, bend to large angles along elliptical and parabolic trajectories while preserving their unique nondiffracting and self-healing nature. Furthermore, we show that the circular NABs found previously represent only a special case of the elliptical MABs, whereas the parabolic WABs represent a perfect counterpart of the Airy beams but without the paraxial limit and are absolutely scalable for acceleration control. Experimentally, we demonstrate finite-energy MABs and WABs bending to large angles beyond the paraxial limit and observe their nondiffracting and self-healing propagation in free space. Such new and generalized NABs provide larger degrees of freedom for launching and controlling the desired beam trajectories for practical applications. This approach is applicable to other linear wave systems in nature, ranging from electromagnetic and elastic waves to matter waves.

Let us first start with MABs by solving the HE in elliptic coordinates. Here we seek a one-dimensional MAB $M(x)$ propagating along an ellipse in the $x$-$z$ plane with the two semi-axes $a$, $b$ of the ellipse oriented along the transverse $x$ and longitudinal $z$ axes, as sketched in Fig. 1(a). By setting $a<b$, the corresponding elliptic coordinates can be constructed via the identities $x=h\sinh\xi\sin\eta$, $z=h\cosh\xi\cos\eta$, where $\xi\in[0,\infty)$ and $\eta\in[0,2\pi)$ are the radial and angular variables, respectively, and $h=|a^2-b^2|^{1/2}$ is the interfocal separation. Assuming that the MAB takes the form of $M(\xi,\eta)=R(\xi)\Theta(\eta)$, the HE can be split into the following modified and canonical Mathieu differential equations

$$\frac{d^2 R(\xi)}{d\xi^2} - (\beta - 2q\cosh 2\xi)R(\xi) = 0 \tag{1a}$$

$$\frac{d^2 \Theta(\eta)}{d\eta^2} + (\beta - 2q\cos 2\eta)\Theta(\eta) = 0 \tag{1b}$$

where $\beta$ is the separation constant, $q=k^2h^2/4$ is a parameter related to the ellipticity of the coordinate system, and $k=2\pi/\lambda$ is the wave number ($\lambda$ is the wavelength in the medium). Solutions of Eqs. (1a) and (1b) are given by radial and angular Mathieu functions [15]. To form a MAB, the beam constructed from those solutions shall travel along an ellipse and preserve the shape with regard to the elliptic coordinate system. For simplicity, we consider a MAB as

$$M(\xi,\eta) = R_m(\xi;q)(ce_m(\eta;q) - ise_m(\eta;q)) \tag{2}$$

where $R_m(\xi;q)$ represents a radial Mathieu function, while $ce_m(\eta;q)$ and $se_m(\eta;q)$ correspond to the even and the odd solutions of angular Mathieu functions at the same order $m$, respectively. Note that Eq. (2) describes the typical optical MABs of our interest, for which $m>(2q)^{1/2}$ is usually a relatively large number so that the even and odd radial Mathieu functions tend to merge [13,15,16].

Similar to the circular NAB cases [10,11], Eq. (2) indicates that a perfect MAB circulates clockwise along an ellipse while preserving its shape in the elliptical coordinate system, thus representing a longitudinal elliptic vortex [16]. However, to construct an ideal MAB with infinite energy, both forward and backward propagating components are required. For a physically realizable beam emitted from a single source and spatially self-accelerating while propagating along the positive $z$ axis, we take a similar approach used in our previous work [11] to introduce a finite energy MAB. By transforming back into the Cartesian coordinates and shifting the main lobe of the beam close to zero, the finite MAB takes the following form at $z=0$

$$M_1(x) = \exp(-\alpha x)H(x+\sqrt{m^2-2q}/k)R_m(\mathrm{Re}(\mathrm{arccosh}(i(x+\sqrt{m^2-2q}/k)/h));q) \tag{3}$$

where $H(x)$ is a Heaviside function, Re means the real part, and $\alpha$ is a positive real number.

Apparently, such a beam contains either a forward or a backward propagating component along an upper half-ellipse with a semi-axis $a \approx (m^2-2q)^{1/2}/k$, resulting in a maximum self-bending of 180°. The Fourier spectrum associated with such a MAB without the exponential truncation can be determined by

$$\Psi_1(f_x) = (k^2 - f_x^2)^{-1/2} \exp(if_x\sqrt{m^2 - 2q}/k + i\arctan(\frac{se_m(\arccos(f_x/k);q)}{ce_m(\arccos(f_x/k);q)})) \qquad (4)$$

where $f_x$ represents the transverse spatial frequency.

Simply by interchanging the $x$ and $z$ coordinates, we can readily construct the MABs for the case $a>b$, for which the beam propagates along the half-ellipses with $a \approx (m^2+2q)^{1/2}/k$. In this case, the beam profiles and their spectra can be determined by the following equations

$$M_2(x) = \exp(-\alpha x) H(x + \sqrt{m^2 + 2q}/k) R_m(\text{arccosh}((x + \sqrt{m^2 + 2q}/k)/h); q) \qquad (5)$$

$$\Psi_2(f_x) = (k^2 - f_x^2)^{-1/2} \exp(i(f_x\sqrt{m^2 + 2q}/k + \text{arccot}(\frac{se_m(\arcsin(f_x/k);q)}{ce_m(\arcsin(f_x/k);q)}))) \qquad (6)$$

Equations (4) and (6) provide the phase mask information for the generation of the MABs from the Fourier space [2,3].

To examine the beam propagation dynamics of the above formulized MABs, we perform numerical simulations by solving the nonparaxial wave equations. The results at $\lambda$=532nm, $m$=80, $h$=8.7μm, $\alpha$=$10^4$, are juxtaposed in Figs. 1(b)-1(f), where (b)-(c) depict the beam profiles, the intensities, and the phases of the Fourier spectra of the MABs at $z$=0 for $a<b$, $a=b$, and $a>b$. Apparently, at $a=b$ ($h$=0), the MAB turns into a circular NAB. The side-view propagations of the three beams are depicted in Figs. 1(d)-1(f), respectively, where all the beams start from a given position on the negative $z$ axis. Note that the horizontal and vertical scales are identical in the figures. Clearly, the constructed MABs self-bend to large angles along elliptical trajectories in contrast to the circular trajectory of a circular NAB as shown in Fig. 1(e). Although the three beams at $z=0$ are quite similar in both real and Fourier spaces (the intensity of the spectra are identical for all three beams), their trajectories of propagation and beam diffraction properties are distinctly different. As one can notice, only the case $a=b$ corresponds to a true nondiffracting NAB, while for the other two cases $a<b$ and $a>b$, the size of the main lobe of the MABs changes slightly after a long propagation. The diffraction property of the MABs will be revisited below.

We point out that Eqs. (4) and (6) contain nontrivial information for unraveling the underlying physics of the MABs. One simple fact is that the beam structure is determined by the combination of $m$ and $q$. This means that, at a certain desired beam width, one can set the beam into different trajectories, representing a larger freedom for beam engineering in comparison to circular NABs. Importantly, under the condition $q \to 0$ or $m \to +\infty$, Eqs. (4) and (6) tend to merge and eventually turn into Eq. (5) in [11] for the circular NABs except for a trivial phase constant, as illustrated in Fig. 1(c). Under such a condition, $a \to b$, and consequently the elliptic trajectory of the beam gradually changes into a circle [see Figs. 1(e)-1(f)]. This clearly indicates that the circular NABs demonstrated recently in [10,11] represent a special case of the MABs presented in this work. In addition, we mention that Eqs. (4) and (6) derived here only involve the solutions of Eq. 1(a), which indicates that one can experimentally construct MABs solely with angular Mathieu functions, without the need of solving the high order modified Mathieu equations.

Next, following a similar procedure for the MABs, let us solve the HE equation for the WABs in parabolic coordinates with $z=\sigma\tau$, $x=(\tau^2-\sigma^2)/2$, $\tau\in(-\infty,+\infty)$ and $\sigma\in[0,\infty)$. Assuming that the WAB takes the form of $W(\sigma,\tau)=\Phi(\sigma)\vartheta(\tau)$, the HE can be split into [14,15]

$$\frac{d^2\Phi(\sigma)}{d\sigma^2}+(k^2\sigma^2+2k\gamma)\Phi(\sigma)=0 \tag{7a}$$

$$\frac{d^2\vartheta(\tau)}{d\tau^2}+(k^2\tau^2-2k\gamma)\vartheta(\tau)=0 \tag{7b}$$

where $2k\gamma$ is the separation constant. The solutions of Eqs. (7a) and (7b) can be determined by the same Weber differential equation (also called parabolic cylinder differential equation) but the corresponding eigenvalues have opposite signs [14,15]. To avoid complex notations, we set $\gamma>0$ and construct a WAB propagating towards the positive $z$ axis as

$$W(\sigma,\gamma) = W_p(\sigma;\gamma)(W_e(\tau;-\gamma) - iW_o(\tau;-\gamma)) \tag{8}$$

where $W_p$ can be either an odd or an even solution of Eq. (7a), while $W_e$ and $W_o$ are the even and odd solutions of Eq. (7b), respectively. The Fourier spectrum associated with such a WAB reads as

$$\varphi(f_x) = \frac{\exp(i\gamma f_x/k + i\gamma \ln(\tan(\arccos(f_x/k)/2)))}{\sqrt{k^2 - f_x^2}} \tag{9}$$

By taking $f_x/k \ll 1$, one can readily prove that Eq. 9 turns into $\exp(-i\gamma f_x^3/3k^3)/k$, representing the typical Fourier spectrum of an Airy beam. This clearly indicates that the Airy beam is the paraxial approximation of our WAB. In addition, we point out that, although such a WAB comes from the exact solution of the HE and has infinite energy, it contains only a forward or a backward propagation component, different from the MABs and circular NABs. Therefore, a physically realizable finite-energy WAB can be established solely by introducing an exponential aperture. By transforming back into the Cartesian coordinates, such a truncated WAB takes the form at $z=0$

$$W(x) = \exp(-\alpha x)W_p(\sqrt{x + \gamma/k};\gamma) \tag{10}$$

From Eq. (9), one can readily deduce the phase information needed for generating the WAB using the Fourier transformation method.

Figure 2 shows a typical example of our WABs at $\lambda=532$nm, $\gamma=40$, and $\alpha=10^4$ and its comparison with an Airy beam with a similar beam size, where (a)-(d) depict the beam profiles, phases of the Fourier spectra [intensity is the same as that in Fig. 1(c)], and side-view propagations of the WAB and the Airy beam, and (e) illustrates the evolution of the beam widths over propagation for different beams. From Figs. 2(c)-2(e), it is obvious that, although the WAB and the Airy beams are all supposed to follow parabolic trajectories, the Airy beam breaks up quickly when bending to large angles. In sharp contrast, the WAB can maintain their fine features well beyond the paraxial regime. Similar to the MABs, although the WABs are shape-preserving solutions in parabolic coordinates, they are no longer perfectly diffraction-free after being transformed back to Cartesian coordinates. Nevertheless, the diffraction of the MABs and WABs are much weaker and largely suppressed in comparison with that of a Gaussian beam of the same size, as shown in Fig. 2(e). Considering there are no true linear nondiffracting beams in reality, it is thus reasonable to consider the MABs and WABs as nondiffracting beams.

Let us now compare the trajectories and accelerations of the WABs and the Airy beams. A WAB determined by Eq. (10) follows $x=-(k/4\gamma)z^2$, while an Airy beam follows $x=-(1/4k^2x_0^3)z^2$ with $x_0$ being the normalization constant for the coordinate $x$. As such, a WAB accelerates along parabolic trajectories determined by the order $\gamma$, similar to an Airy beam. From Eq. 7, one can readily deduce that the WABs are scalable as well. Under the normalization of $x_0$, the beam trajectory turns into $x=-(kx_0/4\gamma)z^2$, and the corresponding beam profile is determined by Eq. (10) with the solutions of Eq. (7a) recasted by normalizing $k$ with $1/x_0$. Thus, one can control the acceleration of WABs simply by squeezing or expanding the transverse coordinate $x$. Apparently, the WABs represent a perfect counterpart of the Airy beams beyond the paraxial limit. Notice that the phase mask embedded in Eq. (9) does not contain any special function. Therefore, many applications proposed for the Airy beams may well be implemented with controllable WABs.

To experimentally demonstrate the MABs and WABs, we use the same setup used previously for generating circular NABs [11], except that now the holographic masks are reconfigured with both the phase and intensity information required to generate the desired accelerating beams. Specifically, a laser beam (λ=532 nm) modulated by a spatial light modulator is sent into a Nikon 60x water-immersion objective lens (NA=1.0, $f$=2.3mm) to synthesize the beams. To directly visualize the beam path, a nanosuspension containing 50nm polystyrene beads is used to scatter the light. Comparing to the conventional method of directly Fourier transforming the phase masks [2-4], our computer-generated holography technique [11,17] is more feasible for constructing the NABs due to the fine features in the phase gradient [see Fig. 1(c) and Fig. 2(b)] and the ability to encode the intensity information [17]. Our experimental results are shown in Fig. 3, where (a)-(e) are the direct top-view photographs if the beam propagation is extrapolated from scattered light, and (f) and (g) display the snapshots of the transverse patterns together with the intensity profiles (along the $x$-direction) taken at $z$=0 and $z$=50μm as marked by the dashed lines in (d, e). From these figures, it can be seen that different elliptical or circular beam paths can be pre-designed by employing different holographic masks according to Eqs. (4) and (6), in agreement with our theoretical analysis of the MABs shown in Fig. 1 but at $m$=3132 and $q$=2.6×10$^6$. The results in Figs. 3(d)-3(g), corresponding to Figs. 2(c) and 2(d), clearly illustrate that the WAB ($\gamma$=1200) preserves its structure much better than the Airy beam does.

Finally, we demonstrate the self-healing property of our MABs and WABs - an interesting feature typical for nondiffracting self-accelerating wave packets [3,4,10]. Figure 4 depicts the numerical and experimental results of the side-view propagation when a MAB or a WAB encounters an obstacle along its trajectory. In these cases, the obstacle partially blocks the accelerating beams around the main lobe. As expected, the MABs and WABs restore their beam structures during subsequent propagation, as if they could overcome the obstacle. The self-healing behaviors of the MABs and WABs indicate the caustic nature of the beam structures, which might provide more controllability from the point of view of the nonparaxial caustics [9,18].

To summarize, we have proposed and demonstrated new families of nonparaxial self-accelerating beams, leading to a novel, complete picture in terms of accelerating beams. In particular, we show that, by designing the MABs and WABs, optical beams with large-angle self-bending, nondiffracting, and self-healing features along designed curved trajectories can be achieved, well beyond the paraxial condition required for the Airy beams. One can directly apply these beams in a variety of applications such as light-induced plasma channels [6], micro-particle manipulations [5] and surface plasmon routing in the general nonparaxial geometry [7]. In addition, the results presented here can be easily extended to other linear wave systems in nature. This approach opens a new door for exploring accelerating wave packets in the nonparaxial regime. Specifically, this raises many intriguing fundamental questions. For instance, could the MABs survive in the presence of nonlinearities [19]? If so, is it possible to realize an optical boomerang traveling along different closed trajectories? Since the results directly stem from the Maxwell equations, will such sharply bending NABs bring about new spin-orbital interaction dynamics [20]? In addition, we expect that many interesting and exciting results are yet to come when this intriguing concept is applied to the time-domain [21] or beyond the diffraction limit [22].

This work was supported by the US ARO MURI program (W911NF-09-1-0539), AFOSR (FA9550-12-1-0111), and the NSF Nano-scale Science and Engineering Center (CMMI-0751621) and PHY-1100842. Y. H. acknowledges support from a MELS FQRNT Fellowship. We thank A. Salandrino for assistance.

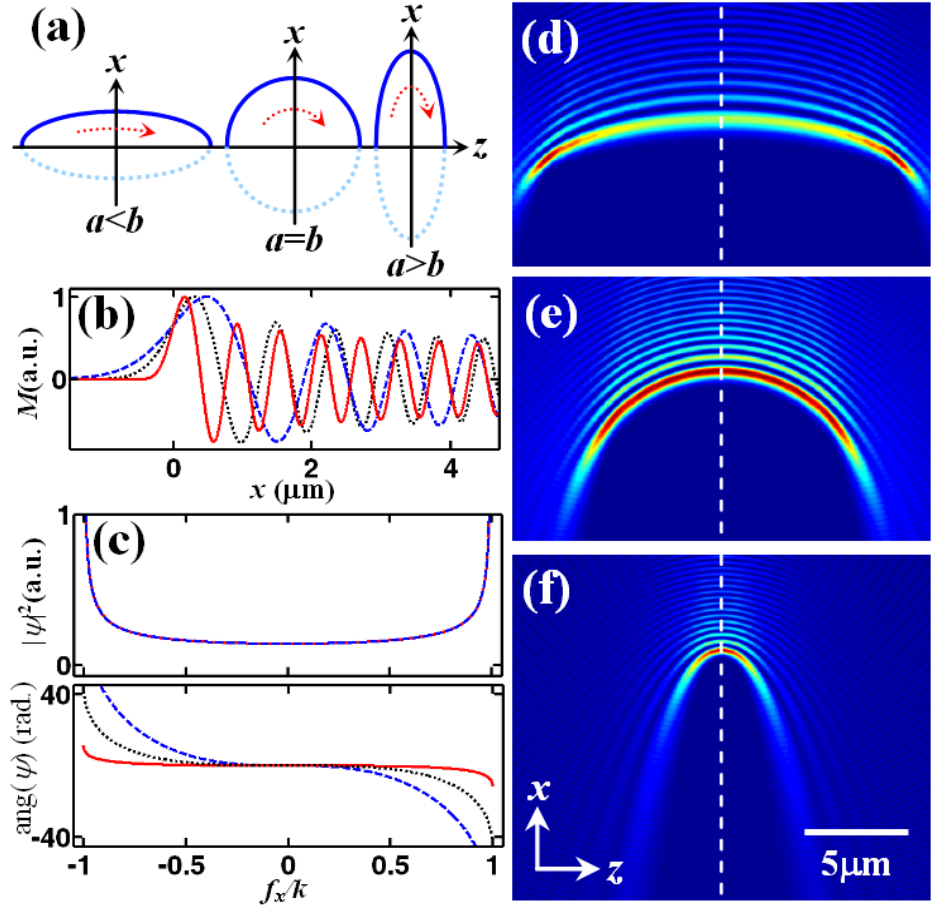

FIG. 1 (color online). (a) Illustrations of different trajectories associated to Mathieu accelerating beams (MABs); (b) Amplitude of the MABs at *a<b* (red, solid), *a=b* (black, dotted), and *a>b* (blue, dashed) at *z*=0; (c) Intensity (top) and phase (bottom) of the Fourier spectra of the MABs in (b) as a function of the normalized spatial frequency ($f_x/k$); (d-f) Side-view propagations of the MABs in (b), where the *z*=0 position is marked by the white dashed line. Note that the horizontal and vertical scales are identical in the figures.

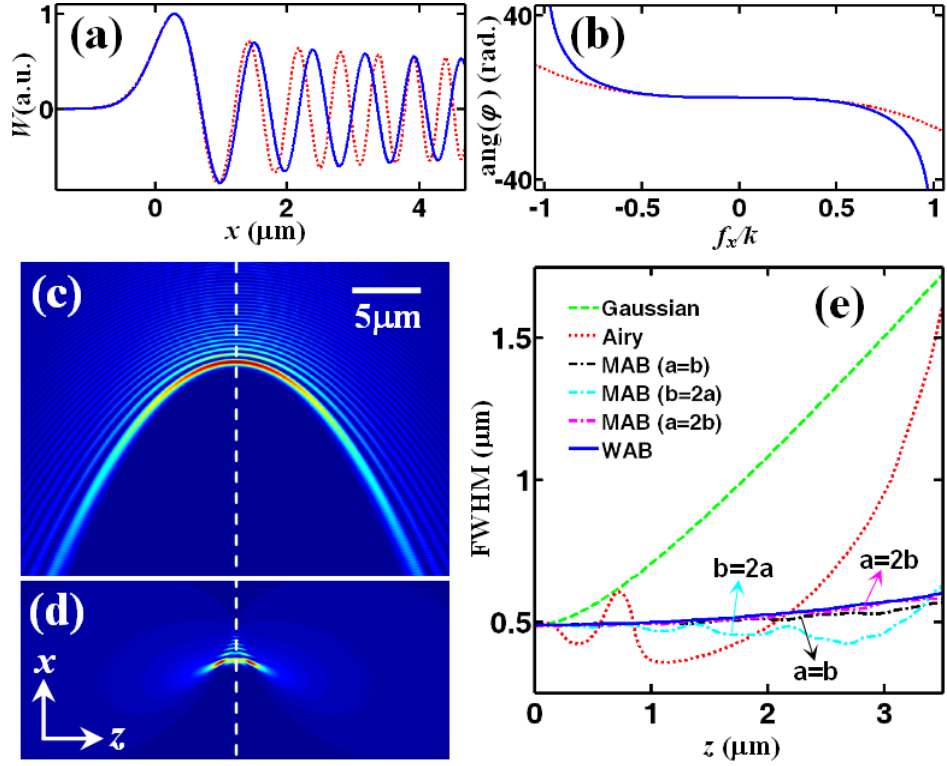

FIG. 2 (color online). Amplitude at $z=0$ (a) and phase of the Fourier spectra (b) of a Weber accelerating beam (WAB) (blue, solid) and an Airy beam (red, dotted); (c) and (d) depict the side-view propagation of the WAB (c) and the Airy beam (d) corresponding to the profiles in (a), where the white dashed line marks the $z=0$ position; (e) illustrates the FWHM of the main lobe of the beam versus the propagation length $z$ for different beams with the same main lobe size at input.

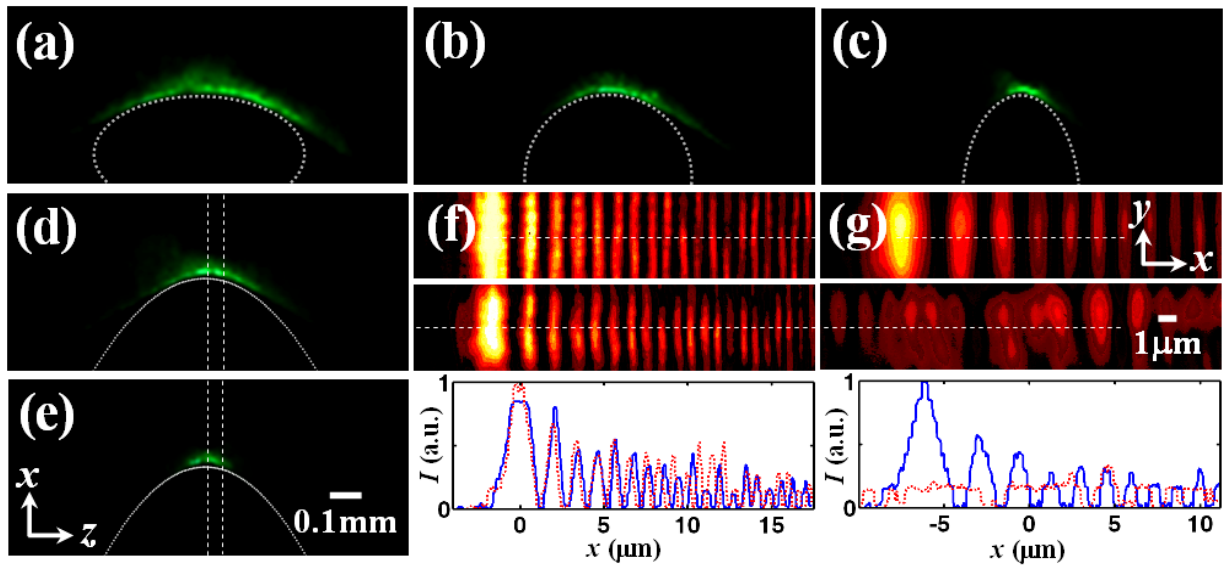

FIG. 3 (color online). Experimental side-view photography of the generated MABs at $a<b$ (a), $a=b$ (b), and $a>b$ (c), a WAB (d) and an Airy beam (e) taken from scattered light, where the dotted curves plot the pre-designed trajectories. (f) and (g) are the transverse intensity distributions recorded at $z=0$ (f) and $z=50\mu m$ (g) as marked by the dashed lines in (d, e), where the top and middle rows correspond to the WAB and Airy beam, respectively, and the bottom depicts the intensity profiles taken along the dashed lines (marked in the transverse intensity patterns shown in the upper two rows). The dashed (red) and solid (blue) lines are corresponding to the Airy and the WAB beams, respectively.

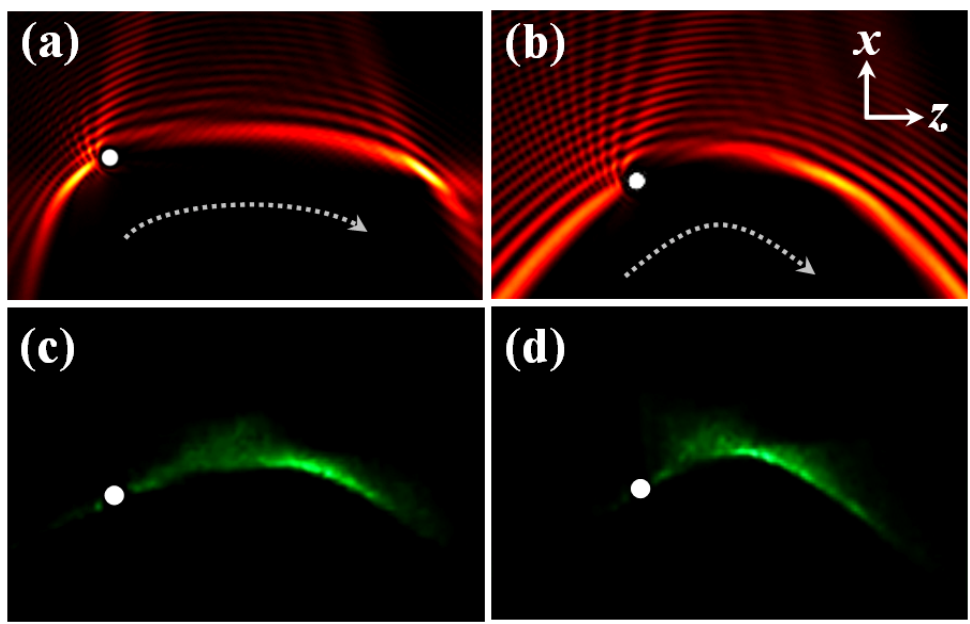

FIG. 4 (color online). Numerical simulations (top) and experimental results (bottom) of the self-healing properties of a MAB (a, c) and a WAB (b, d). (a) and (b) are for beams corresponding to Figs. 1(d) and 2(c), and (c) and (d) are corresponding to Figs. 3(a) and 3(d), respectively. The main lobe of the beam is blocked by an opaque obstacle (white dot) at a certain propagation distance. The diameter of the white dots in (a,b) is 1μm, and that in (c,d) is about 50μm.